\documentclass[11pt]{article}
\usepackage{fullpage,mathptmx,times,amsmath,mathdots,amssymb,epsf,graphicx}
\advance\parskip 4pt

\newtheorem{theorem}{Theorem}[section]
\newtheorem{definition}[theorem]{Definition}

\newtheorem{proposition}[theorem]{Proposition}

\newtheorem{condition}[theorem]{Condition}
\def\proof{\par\kern-\medskipamount\noindent\textbf{Proof.}~~}
\def\remark{\par\kern\medskipamount\noindent\textbf{Remark~\stepcounter{theorem}\arabic{section}.\arabic{theorem}}~}

\catcode`\@ 11
\def\@maketitle{\newpage\null\par\kern-2em%
  \begin{center}%
  \let \footnote \thanks
    {\LARGE \@title \par}%
    \vskip 1.0em%
    {\large
      \lineskip 0.25em%
      \begin{tabular}[t]{c}%
        \@author
      \end{tabular}\par}%
    \vskip 0.5em%
    {\large \@date}%
  \end{center}%
  \par}
\@addtoreset{equation}{section}
\def\overl@ss#1#2{\vcenter{\offinterlineskip
        \ialign{$\m@th#1\hfil##\hfil$\crcr#2\crcr\raise0.6ex\hbox{$<$}\crcr } }}
\def\overgr@ater#1#2{\vcenter{\offinterlineskip
        \ialign{$\m@th#1\hfil##\hfil$\crcr#2\crcr\raise0.6ex\hbox{$>$}\crcr } }}
\def\gl{\mathrel{\mathpalette\overl@ss>}}
\def\lg{\mathrel{\mathpalette\overgr@ater<}}
\long\def\@makecaption#1#2{\vskip\abovecaptionskip
  \sbox\@tempboxa{\small #1: #2}%
  \ifdim \wd\@tempboxa >\hsize \small #1: #2\par
  \else \global \@minipagefalse \hb@xt@\hsize{\hfil\box\@tempboxa\hfil}\fi
  \vskip\belowcaptionskip}
\def\stbox#1{\unskip\hbox to 0.245\textwidth{\small\hfill #1\hfill}}
\def\stext#1#2#3#4{\kern-\smallskipamount
  \hbox to \textwidth{\hss\stbox{#1}\stbox{#2}\stbox{#3}\stbox{#4}\hss}}
\def\sfigs#1#2#3#4{\kern\smallskipamount
  \centerline{\epsfxsize0.245\textwidth\epsfbox{#1.eps}%
  \epsfxsize0.245\textwidth\epsfbox{#2.eps}%
  \epsfxsize0.245\textwidth\epsfbox{#3.eps}%
  \epsfxsize0.245\textwidth\epsfbox{#4.eps}}}

\numberwithin{figure}{section}
\catcode`\@ 12

\let\==\bar
\let\@=\mathbf

\def\circ{\ifmmode\mathchar"220E\else$\mathchar"220E$\fi}
\def\Wr{\mathop{\mathrm{Wr}}\nolimits}
\let\trueint=\int
\let\truesum=\sum
\def\int{\mathop{\textstyle\trueint}\limits}
\def\sum{\mathop{\textstyle\truesum}\limits}

\let\<=\langle
\let\>=\rangle
\def\Real{{\mathbb{R}}}
\def\sech{\mathop{\rm sech}\nolimits}
\def\half{{\textstyle\frac12}}
\def\rank{\mathop{\rm rank}\nolimits}
\let\next=\phi\global\let\phi=\varphi\global\let\varphi=\next
\renewcommand\labelitemi{\ifmmode\circ\else$\circ$\fi}

\begin{document}
\title{\bf Elastic and inelastic line-soliton solutions of the 
Kadomtsev-Petviashvili II equation}
\author{Gino Biondini$^1$ and Sarbarish Chakravarty$^2$\\[1ex]
\small$^1$\it\ 
State University of New York, Department of Mathematics, Buffalo, NY 14260-2900\\
\small$^2$\it\
University of Colorado, Department of Mathematics, Colorado Springs, CO 80933-7150}
\date{}
\maketitle
\begin{abstract}
\noindent
The Kadomtsev-Petviashvili II (KPII) equation admits a large variety of
multi-soliton solutions which exhibit both elastic as well as inelastic
types of interactions. This work investigates a general class 
of multi-solitons which were not previously studied, and which
do not in general conserve the number of line 
solitons after interaction. The incoming and outgoing line solitons  
for these solutions are explicitly characterized by analyzing the 
$\tau$-function generating such solutions. 
A special family of $N$-soliton solutions is also considered in this article.
These solutions are characterized by elastic soliton interactions,
in the sense that amplitude and directions of the individual line solitons
as $y\to\infty$ are the same as those of the individual line solitons
as $y\to-\infty$.
It is shown that the solution space of these elastic $N$-soliton solutions 
can be classified into $(2N-1)!!$ disjoint sectors which are characterized 
in terms of the amplitudes and directions of the $N$ line solitons.  
\end{abstract}

\vglue\bigskipamount
\section{Introduction} 
\label{s:introduction}
\noindent
The purpose of this work is to present a characterization
of a large family of 
real, non-singular, line-soliton solutions of the KPII equation
\begin{equation}
(-4 u_t + u_{xxx} +6u u_x)_x + 3u_{yy} = 0\,
\label{e:KP}
\end{equation}
where $u=u(x,y,t)$ and subscripts $x$, $y$ and $t$ denote partial derivatives.
The KP equation is perhaps the prototypical (2+1)-dimensional integrable 
evolution equation originally derived \cite{SovPhysDoklady15p539}
as a model for small-amplitude, long-wavelength, weakly two-dimensional 
solitary waves in a weakly dispersive medium, and arises in many different 
applications including water waves and plasmas (for a review, 
see e.g. \cite{InfeldRowlands}). The various aspects
related to the integrability of the KP equation have been studied extensively, 
and a large number of exact solutions have been found. These works
are documented in several monographs (see e.g., 
Refs.~\cite{AblowitzClarkson,InfeldRowlands,NMPZ1984} and references therein). 
There are two versions
of the KP equation depending on the sign of the dispersion namely, KPI (for positive
dispersion) and KPII (for negative dispersion). Here we consider the KPII equation.
 
Among the exact solutions of KPII, perhaps one of the most well known class 
of real, non-singular solutions is the line-soliton solutions. The simplest 
type is the one-soliton
solution which is a traveling wave in $xy$-plane, and is localized along a
line. A straightforward generalization of this solution to a multi-soliton 
configuration of $N$ line solitons was also found in earlier 
works~\cite{JPSJ1976v40p286,PLA95p1}.
In the generic case, these $N$-soliton solutions form a pattern of $N$ intersecting 
straight lines in the $xy$-plane apart from small spatial shifts arising from the 
pairwise interactions between any two lines.
We refer to these solutions as the {\em ordinary} $N$-soliton solutions
which can be parametrized by $2N$ parameters namely, amplitude and 
direction (in the $xy$-plane) of the $N$ line solitons.
However, it has been shown theoretically and experimentally that
it is not possible to obtain ordinary $N$-soliton solutions for {\em all}
choices of the soliton parameters (see e.g., Ref.~\cite{InfeldRowlands}).
That is, the parameter space of ordinary $N$-soliton solutions is {\em not}
simply the $N$-fold Cartesian product of the parameter space of one-soliton 
solution of the KPII equation. Subsequent work has revealed that
other families of line soliton solutions exist in addition to the ordinary 
$N$-soliton solutions.
The simplest of these solutions describes the so called ``Y-junction'' solutions,
which describes
resonant interaction of two line solitons, and has been known since 1977 
\cite{JFM1977v79p171,PRL1977v38p377}. 
In this case, the three solitons with wave-numbers and frequencies 
$({\bf k}_a, \omega_a),\, a = 1,2,3$,
satisfy the three-wave resonance conditions:\, 
${\bf k}_1+{\bf k}_2={\bf k}_3$ and $\omega_1+\omega_2=\omega_3$.
More general resonant solutions have also been obtained in 
Refs.~\cite{medina,JPSJ1983v52p749,pashaev}. Recently, in Ref.~\cite{jphysa36p10519},
the authors found a large family of soliton solutions in which
an arbitrary number $N_-$ of incoming line solitons interact resonantly 
via intermediate line solitons which produce web like patterns in the 
$xy$-plane, and then form an arbitrary number $N_+$ of outgoing line solitons, 
where $N_- \neq N_+$, in general. 
We refer to these multi-solitons solutions as the
$(N_-,N_+)$-soliton solutions of KPII.
In particular, the case $N_-=N_+=N$ yields yet 
another kind of $N$-soliton solutions
which however differ significantly from the ordinary $N$-soliton solutions in
their interaction patterns. The current work is motivated by these
recent studies which indicate that the solitonic sector of the KPII equation is 
richer than previously thought as many other families of line-soliton 
solutions exist in addition to the ordinary $N$-soliton solutions.

In this article we are primarily concerned with the characterization
of the incoming and outgoing line solitons of a generic multi-soliton
configuration of KPII as well as the classification of a particular class 
of multi-soliton solutions called the \textit{elastic} $N$-soliton solutions.
The paper is organized as follows. In section~\ref{s:soliton}, we discuss
the asymptotic behavior of the $\tau$-function underlying the multi-soliton
solutions of KPII. In particular, we show that as $y \to \pm \infty$,
the solution $u(x,y,t)$ decays exponentially in the $xy$-plane except along 
certain rays which correspond to the incoming and outgoing 
line solitons. We then characterize these asymptotic structures in terms of
the parameters of the $\tau$-function. In section \ref{s:elastic}, we 
study a special class of solutions called the elastic $N$-soliton solutions.
We show that these solutions can be classified into $(2N-1)!!$ inequivalent
types determined by the $2N$ soliton parameters comprising of $N$ pairs of
amplitudes and directions associated with the incoming or outgoing line solitons.
Moreover, from a given set of admissible soliton parameters
one can explicitly determine an equivalence class of elastic $N$-soliton solutions
such that any two solutions in the equivalence class exhibit similar interaction
patterns and have the same set of incoming and outgoing line solitons. 
In this paper, we only present the results and discuss some of the important features
of the multi-soliton solutions of KPII. The proofs of these results and more 
detailed discussions can be found in Refs.~\cite{GBSC,GBSCYK}.

We point out that the elastic $N$-soliton solutions of KPII equation
was also recently addressed in Ref.~\cite{Kodama} where an elegant
characterization of these solutions were presented in terms of the
Schubert cell decomposition of the Grassmann variety Gr($N,2N$).
Here we follow a different approach motivated by the physical problem
of identifying the distinct types of elastic $N$-soliton solutions with
the the corresponding parameter space of soliton amplitudes and velocities.
Finally, we note that line soliton solutions with novel web like
spatial structures were also found 
recently in several other $2+1$-dimensional integrable equations. 
Examples include Refs.~\cite{jphysa35p6893,jphysa36p9533} for a coupled KP system,
and Ref.~\cite{jphysa37p11819} where similar solutions were found in
discrete integrable systems such as the two-dimensional Toda lattice
and its fully- and ultra-discrete analogues. 

\section{Asymptotic line solitons}
\label{s:soliton}
\noindent
In this section we investigate the line solitons of 
the KPII equation and the asymptotic properties of the $\tau$-functions
generating such solutions.
The solution $u(x,y,t)$ of the KPII equation can be obtained from 
the $\tau$-function $\tau(x,y,t)$ via the relation
\begin{equation}
u(x,y,t)= 2(\log\tau(x,y,t))_{xx}\,.
\label{e:u}
i\end{equation}
It is well-known (see, e.g. Refs.~\cite{PLA95p1,Hirota,MatveevSalle})
that $\tau(x,y,t)$ can be expressed in the form of a Wronskian 
\begin{equation}
\tau(x,y,t)= \Wr(f_1,\dots,f_N)= \det\begin{pmatrix}
  f_1 &f_2 &\dots &f_N\\
  f_1' &f_2' &\dots &f_N'\\
  \vdots &\vdots & &\vdots\\
  f_1^{(N-1)} &f_2^{(N-1)} &&f_N^{(N-1)}
  \end{pmatrix}\,,
\label{e:tau}
\end{equation}
where $f_n^{(j)}= \partial^j f_n/\partial x^j$, and 
where the functions $\{f_n\}_{n=1}^N$
form a set of linearly independent solutions of the linear system
\[
f_y= f_{xx}\,,\qquad f_t= f_{xxx}\,.
\]
A general family of multi-soliton solutions can be constructed in a simple
way from Eq.~\eqref{e:tau} by choosing each function
$f_n(x,y,t)$ to be a linear combination of real exponentials. That is,
\begin{equation}
f_n(x,y,t)= \sum_{m=1}^{M} a_{nm}\,e^{\theta_m}\,, \quad
n = 1,2, \ldots, N\,,
\label{e:f}
\end{equation}
where, $\theta_m = k_mx+k_m^2y+k_m^3t+\theta_{0m}\,$ for $m=1,\ldots,M$ 
are $M$ phases with real phase parameters $k_1, \ldots, k_M$
and real constants $\theta_{01},\dots,\theta_{0M}$, 
and where the constant coefficients $a_{nm}$ define the $N \times M$
\textit{coefficient matrix} $A:= (a_{nm})$. 
Note that one can naturally identify each $f_n$
with the $n^{\mathrm {th}}$ row and each phase~$\theta_m$ with
 the $m^{\mathrm {th}}$ column of the coefficient matrix $A$, and vice versa. 
Upon substituting
Eq.~\eqref{e:f} into the Wronskian of Eq.~\eqref{e:tau} and then
using the Binet-Cauchy formula to expand the resulting determinant,
we obtain the following explicit form of the $\tau$-function
\begin{equation}
\tau(x,y,t)= \sum_{1\le m_1<\dots<m_N\le M}
  A(m_1,\dots,m_N) \,\,
\exp[\,\theta(m_1,\dots,m_N) \,]
\!\!\prod_{1\le s < r\le N}(k_{m_{r}}-k_{m_{s}}) \,,
\label{e:tauexp}
\end{equation}
where $A(m_1,\dots,m_N)$ is the $N\times N$ minor of $A$
obtained by selecting the columns $1\le m_1<\dots<m_N\le M$,
and $\theta(m_1,\dots,m_N) := \theta_{m_1}+\ldots+\theta_{m_N}$
is a phase combination of $N$ (out of $M$) distinct phases.
The $\tau$-function given above could in general, vanish at 
points $(x,y,t) \in \Real^3$ where the solution $u(x,y,t)$ in 
Eq.~\eqref{e:u} would then have singularities. However, the
following restrictions on the phase parameters $\{k_n\}_{n=1}^M$ 
and on the coefficient matrix $A$ 
are sufficient to guarantee that the resulting
solutions $u(x,y,t)$ of the KPII equation are nonsingular. 
\begin{condition}~(Positive definiteness)
\label{c:positive}
\begin{enumerate}
\vspace{-0.1 in}
\item[(a)]\, The phase parameters are distinct. Hence, without loss
of generality, they can be ordered as $k_1<k_2<\ldots<k_M$.
\vspace{-0.1 in}
\item [(b)]\, The $N \times M$ coefficient matrix $A$ satisfies
$\rank(A) = N$, and $M > N$.
\vspace{-0.1 in}
\item[(c)]\, All non-zero $N \times N$ minors of $A$ are positive.
\end{enumerate}
\end{condition}
From Condition~\ref{c:positive} it is clear
that the coefficient of each exponential term of 
the sum in Eq.~\eqref{e:tauexp} is positive because the 
phase parameters $k_1,\dots,k_M$ are well-ordered and all the 
minors $A(m_1,\ldots,m_N)$ of $A$ are also nonnegative. As a result, 
$\tau(x,y,t)$ is a nonvanishing, positive function for 
all $(x,y,t)\in\Real^3$, and generates a nonsingular solution of the 
KPII equation via Eq.~\eqref{e:u}. If $M < N$, $\tau(x,y,t)=0$
because the set of functions $\{f_n\}_{n=1}^M$ in Eq.~\eqref{e:tau}
is linearly dependent; also when $M=N$, $\tau(x,y,t)$ in 
Eq.~\eqref{e:tauexp} contains only one exponential term which
leads to the trivial solution $u(x,y,t) =0$ in Eq.~\eqref{e:u}.
Therefore, for nontrivial solutions we must have $M>N$ when
there are more than one exponential term in the sum of
Eq.~\eqref{e:tauexp}.

The simplest example is a one-soliton solution
obtained by choosing $N=1, \, M=2$ and 
$f(x,y,t)= e^{\theta_1}+e^{\theta_2}$, with $k_1 < k_2$.
This choice yields the following traveling-wave solution
\begin{equation}
u(x,y,t)= 
\half(k_2-k_1)^2\sech^2\half(\theta_2-\theta_1)= 
\Phi(\@k\cdot\@x+\omega t) \,,
\label{e:onesoliton}
\end{equation}
where $\@x=(x,y)$ and where the wave vector 
$\@k:=(l_x,l_y)=(k_1-k_2,k_1^2-k_2^2)$ and the frequency 
$\omega:=k_1^3-k_2^3$ satisfy the nonlinear dispersion relation
\begin{equation}
-4\omega l_x+l_x^4+3l_y^2=0\,.
\label{e:dispersionrelation}
\end{equation}
For fixed $t$, the solution $u(x,y,t)$ decays exponentially 
in the $xy$-plane except along the line $\theta_1=\theta_2$
whose normal has a slope $c=l_y/l_x = k_1+k_2$.
Such solitary wave solutions of the KPII equation are called
\textit{line solitons}.
Apart from a trivial constant $\theta_{1,0}-\theta_{2,0}$
in Eq.~\eqref{e:onesoliton} corresponding to an overall translation,
a line soliton of KP is characterized by the phase parameters
$k_1, k_2$, or by two physical parameters, namely,
the \textit{soliton amplitude}~$a:=k_2-k_1$
and the \textit{soliton direction}~$c:=k_1+k_2$.

When $c=0$ (i.e., $k_1 = -k_2$), the solution
in Eq.~\eqref{e:onesoliton} becomes $y$-independent and reduces
to the one-soliton solution of the Korteweg-de~Vries (KdV) equation.
However, due to the dependence on the additional spatial variable $y$, 
the multi-soliton solution space of the KPII equation
is much richer than that of the KdV. Indeed we find that 
Eq.~\eqref{e:f} with the
coefficient matrix $A$ satisfying Condition~\ref{c:positive} 
leads to a large class of multi-soliton configurations which
consist of $N_-$ incoming (i.e., as $y \rightarrow -\infty$)
and $N_+$ outgoing (i.e., as $y \rightarrow \infty$)
line solitons. The amplitudes, directions and the number of incoming 
line solitons are in general different from those of the outgoing 
line solitons. In order to characterize
the incoming and outgoing line solitons associated with these solutions,
it is necessary to examine the asymptotic behavior of the $\tau$-function 
in the $xy$-plane as $|y| \to \infty$, and for finite $t$. 
Recall from Eq.~\eqref{e:tauexp}
that $\tau(x,y,t)$ is a linear combination of exponential 
phase combinations
with positive coefficients. The leading order
behavior of the $\tau$-function as $y \to \pm \infty$ in a given 
asymptotic sector of the $xy$-plane is governed by that exponential term
which is dominant in that region. The solution $u(x,y,t)$ generated by the
$\tau$-function is exponentially small at all points
in the interior of any dominant region, and is localized only at
the boundaries of the dominant regions, where a balance exists
between two or more dominant phase combinations in the $\tau$-function 
of Eq.~\eqref{e:tauexp}. The asymptotic properties of the $tau$-function 
and the solution $u(x,y,t)$ can be derived from a systematic analysis of 
these dominant exponential phases. These are summarized below.
\begin{proposition}
\label{P:soliton}
For finite values of $t$, and for generic values of phase parameters
$k_1,\ldots,k_M$, the incoming and outgoing line solitons of 
the $(N_-,N_+)$-soliton solutions of KPII are characterized as follows. 
\begin{enumerate}
\item
As $y\to\pm\infty$, the dominant phase combinations of the
$\tau$-function in adjacent regions of the $xy$-plane contain
$N-1$~common phases and differ by only a single phase.
The transition between any two such dominant phase combinations
$\theta(i,m_2,\dots,m_N)$ and $\theta(j,m_2,\dots,m_N)$
occurs along the line defined by $L_{ij}:\theta_i=\theta_j$,
where a single phase $\theta_i$ in the dominant phase combination
is replaced by a phase $\theta_j$.
\item 
Along the single-phase transition line $L_{ij}$,
the asymptotic behavior of the $\tau$-function as $y \to \pm \infty$ is\break 
determined by
the balance between the two dominant phase combinations
$\theta(i,m_2,\dots,m_N)$ and\break $\theta(j,m_2,\dots,m_N)$ 
in Eq.~\eqref{e:tauexp}, and is given by
\begin{equation*}
\tau(x,y,t)\sim
  C_i\,e^{\theta(i,m_2,\dots,m_N)} + C_j\,e^{\theta(j,m_2,\dots,m_N)}\,.
\end{equation*}
The coefficients $C_i$ and $C_j$ above depend on the phase parameters
$k_i,k_j,k_{m_2},\ldots,k_{m_N}$, and on the $N \times N$ minors
$A(i,m_2,\dots,m_N)$ and $A(j,m_2,\dots,m_N)$ of the coefficient matrix~$A$.
The asymptotic behavior of the solution in a neighborhood 
of a single-phase transition is then obtained from Eq.~\eqref{e:u} as
\begin{equation}
u(x,y,t) \sim
  \half(k_i-k_j)^2\sech^2\big[\half(\theta_i-\theta_j)\big]\,,
\label{e:asympsol}
\end{equation}
which is a traveling wave satisfying the
dispersion relation in Eq.~\eqref{e:dispersionrelation}.
Equation~\eqref{e:asympsol}
has the same form as the one-soliton solution in Eq.~\eqref{e:onesoliton},
and thus it defines an {\em asymptotic} (incoming or outgoing) line soliton
associated with the single-phase transition $i \to j$.
For each asymptotic line soliton,
the soliton amplitude is given by $a_{ij}=|k_i-k_j|$,
and the soliton direction is given by $c_{ij}= k_i+k_j$, which
the direction (slope of the normal vector) of the transition line $L_{ij}$.
\end{enumerate}
\end{proposition}
\noindent
We can label each asymptotic line soliton associated with the
single-phase transition $i \to j$, by the index pair $[i,j]$ which
uniquely identifies the phase parameters~$k_i$ and $k_j$
in the ordered set $\{k_1,\dots,k_M\}$.

\begin{figure}[t!]
\newdimen\figwidth \figwidth 0.375\textwidth
\centerline{\raise0.5ex\hbox{\epsfxsize0.965\figwidth\epsfbox{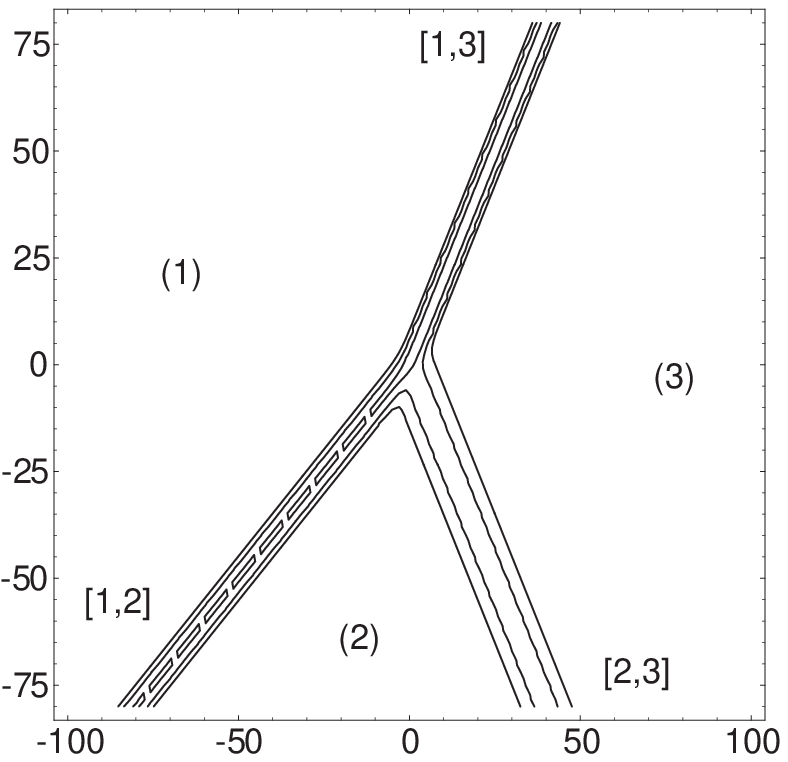}}\quad
\epsfxsize\figwidth\hbox{\epsfbox{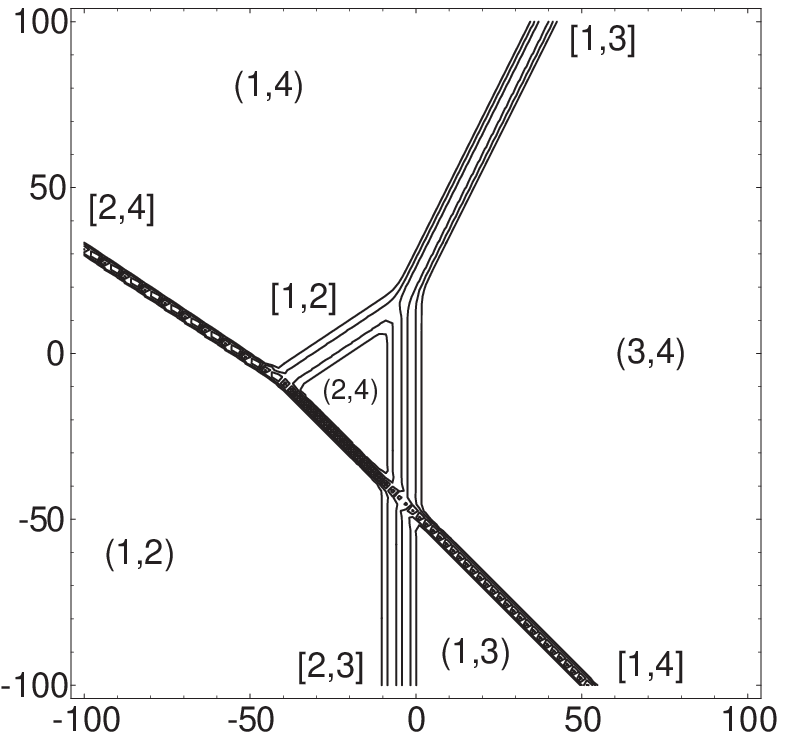}}}
\caption{Dominant phase combinations in the different regions
 of the $x,y$-plane (labeled by the indices in parentheses)
 and the asymptotic line solitons (labeled by the indices in square braces)
 for two different line soliton solutions:
(a)~a Miles resonant solution (Y-junction) with 
   $(k_1,k_2,k_3)=(-\frac32,-\frac12,\frac12,2)$
   at $t=0$;
(b)~an inelastic $(2,2)$-soliton with $(k_1,\dots,k_4)=(-\frac32,-\frac12,0,1)$
 at $t=0$.
 Here and in all of the following figures,
 the horizontal and vertical axes are respectively $x$~and~$y$, and
 the graphs show contour lines of $\log u(x,y,t)$ at a fixed value of~$t$.}
\label{f:dominantphases}
\end{figure}

The simplest instance of a transition of dominant
phase combinations arises for the one-soliton
solution~\eqref{e:onesoliton},
which is localized in the $xy$-plane along the line 
$\theta_1=\theta_2$ defining the boundary of the two half-planes 
where each of the two phases $\theta_1$ and $\theta_2$ dominates.
In the general case, the dominant regions are more complicated 
although the solution $u(x,y,t)$ is still localized along the boundaries 
of these regions. 
For example, Fig.~\ref{f:dominantphases}(a) illustrates a 
$(2,1)$-soliton solution mentioned in the introduction as
a Y-junction \cite{JFM1977v79p171} describing the resonant interaction 
of two line solitons.
This solution corresponds to $N=1,\, M=3$, and is generated by the 
$\tau$-function $\tau(x,y,t)= e^{\theta_1}+e^{\theta_2}+e^{\theta_3}$.
In this case, the $xy$-plane is partitioned into three dominant regions
corresponding to each of the dominant phases $\theta_1$, $\theta_2$
and~$\theta_3$. Once again,
the solution $u(x,y,t)$ is exponentially small in the interior of each
dominant regions, and is localized along the phase transition boundaries:
$\theta_1=\theta_2$,  $\theta_1=\theta_3$ and $\theta_2=\theta_3$.
Each of the asymptotic line solitons labeled by the index pairs [1,2], 
[2,3] and [1,3] are given by Eq.~\eqref{e:asympsol} and satisfy the
one-soliton dispersion relation Eq.~\eqref{e:dispersionrelation}.
While some of the dominant regions have infinite extensions
in the $xy$-plane, others can be bounded, as in the case of the
$(2,2)$-soliton shown in Fig.~\ref{f:dominantphases}(b).
This solution is generated by the $\tau$-function in Eq.~\eqref{e:tau}
with $f_1 = e^{\theta_1}-e^{\theta_4}$ and 
$f_2 = e^{\theta_2}+e^{\theta_3}+ e^{\theta_4}$, i.e., $N=2, \, M=4$.
In addition to the unbounded dominant regions corresponding to
the phase combinations $\theta(1,2), \, \theta(1,4),\, \theta(3,4)$ and 
$\theta(1,3)$, in this case there is also a bounded region in the 
$xy$-plane where $\theta(2,4)$ is the dominant phase combination. 
The boundaries of this
region is formed by the incoming asymptotic line solitons [1,4] and [2,3], 
together with the intermediate line soliton [1,2]. 
Note that the outgoing line solitons [1,3] and [2,4] interact resonantly 
via two Y-junctions, while the incoming soliton pair interact non-resonantly.
Note also that the intermediate segment is a line soliton in its own right,
since the solution is locally given by Eq.~\eqref{e:asympsol},
with $[i,j]=[2,3]$.

According to Proposition \ref{P:soliton}, an asymptotic line
soliton corresponds to a dominant balance between two phase
combinations in the $\tau$-function. But one still
needs to identify which particular phase combinations are indeed dominant 
in a given $\tau$-function as $|y| \to\infty$.
This requires a closer examination at the structure of the $N \times M$ 
coefficient matrix $A$ associated with the $\tau$-function. 
Note that elementary row operations on $A$ given
by $A \rightarrow A'=G\,A$ where $G\in \mathrm{GL}(N,\Real)$,
amounts to an overall rescaling of the $\tau$-function in 
Eq.~\eqref{e:tauexp}, i.e., $\tau(x,y,t)\to \tau'(x,y,t)=\det(G)\,\tau(x,y,t)$.
Since such rescaling leaves the 
solution $u(x,y,t)$ in Eq. \eqref{e:u} invariant, it is possible to choose the 
coefficient matrix $A$ in reduced row-echelon form (RREF)
making use of Gaussian elimination.
Recall that, for an $N\times M$ matrix in RREF, 
the leftmost nonvanishing entry in each nonzero row is called a pivot,
which is normalized to 1, so that the pivot columns are the elements 
of the canonical basis of $\Real^N$.
Throughout the rest of this work we will consider the 
coefficient matrix~$A$ to be in RREF, and 
to satisfy Condition~\ref{c:positive}
and the following additional conditions:
\begin{condition} (Irreducibility)~
\label{c:irreducible}
Each column of $A$ contains at least one nonzero element, and
each row of $A$ contains at least one nonzero element in addition to the pivot.
\end{condition}
Then a detailed analysis of the coefficient matrix $A$ satisfying 
Conditions \ref{c:positive} and \ref{c:irreducible} leads to an
explicit identification of those $i \to j$ single-phase transitions 
which actually occurs as $|y| \to \infty$, for any given $\tau$-function
of Eq.~\eqref{e:tauexp}. As a result, each asymptotic line-soliton 
$[i,j]$ is also explicitly determined by the coefficient matrix $A$.
as given below. Specifically, we have the following results.
\begin{proposition}
\label{P:pairing}
Each index pair $[i,j]$ labeling an asymptotic line soliton of a
$(N_-,N_+)$-soliton solution are uniquely identified with a pair of 
columns of the associated coefficient matrix $A$, as prescribed below.
\begin{enumerate}
\item
An asymptotic line soliton as $y\to\infty$ is identified by 
a unique index pair $[e_n,j_n]$ with $e_n < j_n$ and where 
$\{e_n\}_{n=1}^N$ label the pivot columns of $A$.
Similarly, an asymptotic line soliton as $y\to -\infty$ is identified
with a unique index pair $[i_n,g_n]$ with $i_n < g_n$ and where
$\{g_n\}_{n=1}^{M-N}$ label the non-pivot columns of $A$.
Thus, the $(N_-,N_+)$-line soliton
solution of KPII generated from the $\tau$-function 
in Eq.~\eqref{e:tauexp} has exactly $N_+= N$ asymptotic line solitons
as $y\to\infty$ and $N_-=M-N$ asymptotic line solitons as $y\to-\infty$.
\item
The necessary and sufficient conditions for an index pair $[i,j]$ 
to identify an asymptotic line soliton
is determined by considering the ranks of two sub-matrices $X_{ij}$ and 
$Y_{ij}$ of $A$. They can be denoted by their column indices as follows:
\begin{equation*}
X_{ij} = \left[ 1,2,\ldots, i-1, j+1, \ldots, M \right] \qquad
Y_{ij} = \left[i+1, \ldots j-1 \right] \,.
\end{equation*}
That is, $X_{ij}$ consists of all consecutive columns to the left of the
i$^{\mathrm {th}}$ column and all consecutive columns to the right of the
j$^{\mathrm {th}}$ column of $A$, while $Y_{ij}$ consists of all consecutive
columns in between the i$^{\mathrm {th}}$ and j$^{\mathrm {th}}$ column of $A$.
The rank conditions are then stated as follows.
\begin{enumerate}
\item
$[i,j]$ identifies an asymptotic line soliton as $y\to\infty$ if and only if\,
$\rank(X_{ij}):=r \le N-1$\, and \break
$\rank(X_{ij}|i)= \rank(X_{ij}|j)= \rank(X_{ij}|i,j)= r+1$.
\item
$[i,j]$ identifies an asymptotic line soliton as $y\to-\infty$ if and only if \,
$\rank(Y_{ij}) := s\le N-1$\, and \break
$\rank(Y_{ij}|i) = \rank(Y_{ij}|j)= \rank(Y_{ij}|i,j)= s+1$.
\end{enumerate}
Above, $(Z|m,n)$ denotes the sub-matrix~$Z$ of $A$ augmented by the 
columns $m$ and $n$ of $A$.
\end{enumerate}
\end{proposition}
\noindent
Note that for the asymptotic line soliton $[e_n,j_n]$ as $y\to\infty$
in Proposition~\ref{P:pairing}(i), $e_n$ is a pivot index but the
index $j_n$ can be either a pivot or a non-pivot index.
Similarly, for the asymptotic line solitons $[i_n,g_n]$ as 
$y\to-\infty$, the index $g_n$ is an non-pivot index, while $i_n$ can be 
either a pivot or a non-pivot index.  
The necessary and sufficient rank conditions in
Proposition~\ref{P:pairing}(ii) provides a constructive method to 
identify the asymptotic line solitons as $y\to\pm\infty$ 
from a given coefficient matrix~$A$ in RREF.
We illustrate these statements by the examples below. 

\noindent
{\em Example 1}:\, Consider the $\tau$-function in Eq.~\eqref{e:tauexp}
with $N=2$ and $M=5$ generated by the coefficient matrix
\begin{equation}
A= \begin{pmatrix}
1 &1 &0 &\!-1 &\!-2\\ 0 &0 &1 &1 &1
\end{pmatrix}
\label{e:3to2}
\end{equation}
The pivot columns of~$A$ are labeled by the indices $\{e_1,e_2\}=\{1,3\}$,
and the non-pivot columns by the indices $\{g_1,g_2,g_3\}=\{2,4,5\}$.
Thus, from Proposition~\ref{P:pairing}(i) we know that there will 
be $N_+=N=2$ asymptotic line solitons as $y\to\infty$, identified by the 
index pairs $[1,j_1]$ and $[3,j_2]$ for some $j_1>1$ and $j_2>3$,
and that there will be
$N_-=M-N=3$ asymptotic line solitons as $y\to-\infty$, identified by the
index pairs $[i_1,2]$, $[i_2,4]$ and $[i_3,5]$, for some
$i_1<2$, $i_2<4$ and $i_3<5$.
We first determine the asymptotic line solitons as $y\to\infty$
using the rank conditions prescribed in Proposition~\ref{P:pairing}(ii).
For the first pivot column, $e_1=1$, we start with $j=2$ and consider
the sub-matrix
$X_{12}= \bigl(\begin{smallmatrix} 0 &\!-1 &\!-2\\ 1 &1 &1
\end{smallmatrix}\bigr)\,$.
Since $\rank(X_{12})=2$ which is greater than $N-1=1$, we conclude 
that the pair $[1,2]$ cannot identify an asymptotic line soliton as 
$y\to\infty$. Incrementing $j$ to $j=3,4,5$ and checking the rank of 
each sub-matrix $X_{1j}$ we find that the rank conditions in 
Proposition~\ref{P:pairing}(ii) are satisfied when $j=4$:
$X_{14}= \bigl(\begin{smallmatrix}\!-2\\1\end{smallmatrix}\bigr)$,
so $\rank(X_{14})=1$ and $\rank(X_{14}|1)= \rank(X_{14}|4)= 
\rank(X_{14}|1,4) =2$. Thus, the first asymptotic line soliton 
as $y\to\infty$ is identified by the index pair $[1,4]$.
For the second pivot, $e_2=3$, proceeding in a similar manner we 
find that $j=4$ does not satisfy the rank conditions (since 
$X_{34}$ has rank~2) but $j=5$ does: 
$X_{35}= \bigl(\begin{smallmatrix} 0 &\!-1 &\!-2\\ 1 &1 &1
\end{smallmatrix}\bigr)\,$,
which yields $\rank(X_{35})=1$ and
$\rank(X_{35}|3)= \rank(X_{35}|5)= \rank(X_{35}|3,5)=2$.
Therefore, the asymptotic line solitons as $y\to\infty$ are given by
the index pairs $[1,4]$ and $[3,5]$.

\begin{figure}[t!]
\newdimen\figwidth \figwidth 0.375\textwidth
\centerline{\epsfxsize0.9675\figwidth\raise1.5ex\hbox{\epsfbox{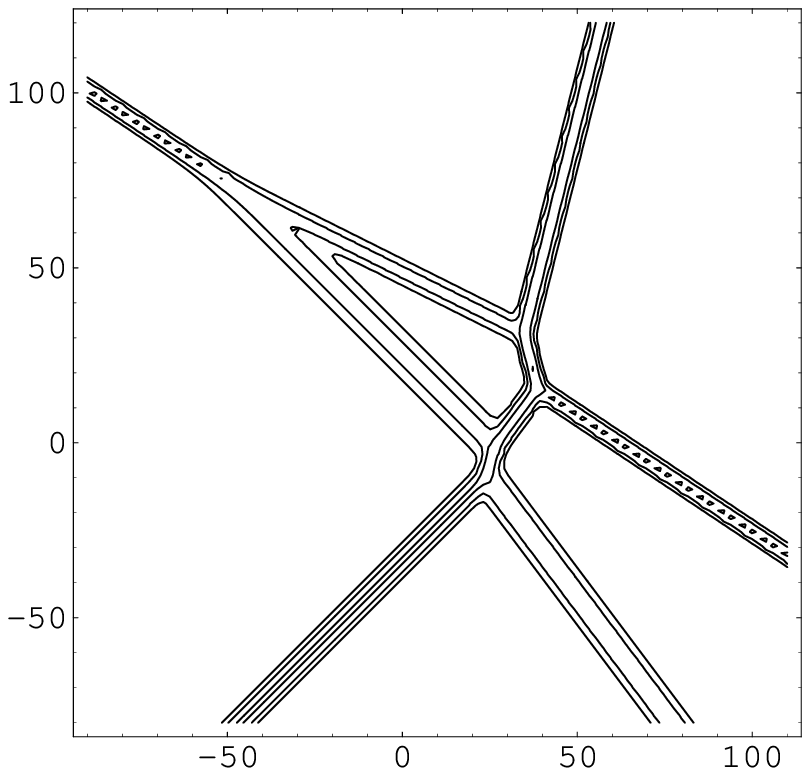}}
\quad
\epsfxsize\figwidth\raise1ex\hbox{\epsfbox{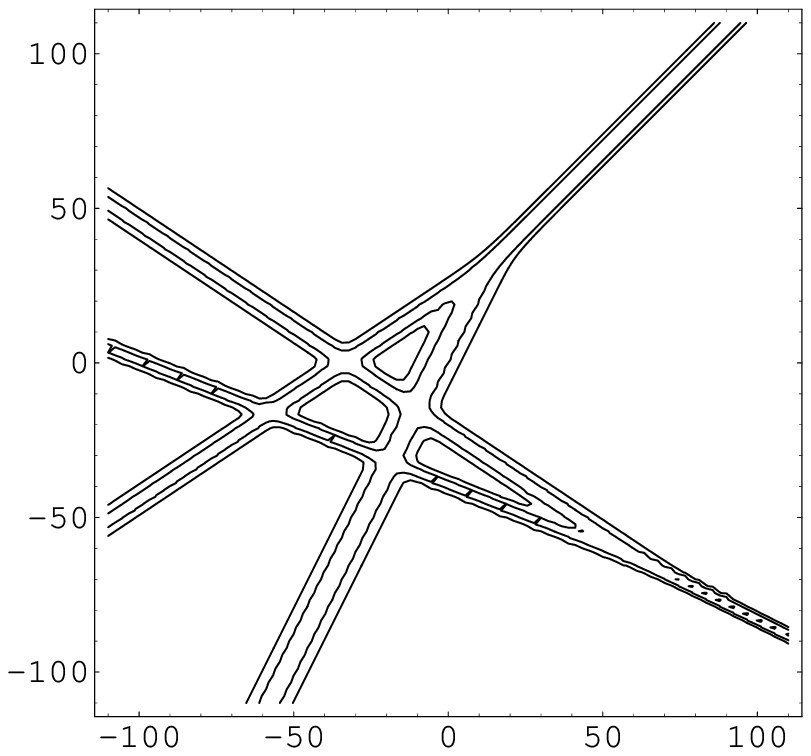}}}
\caption{Line soliton solutions of KPII:
(a)~the (3,2)-soliton solution generated by the coefficient matrix~$A$
in Eq.~\eqref{e:3to2} with
$(k_1,\dots,k_5)=(-1, 0, \frac14, \frac34, \frac54)$ at $t=-32$;
(b)~the (3,3)-soliton solution
generated by the coefficient matrix~$A$ in Eq.~\eqref{e:inelastic3s}
with $(k_1,\dots,k_6)=(-1, -\frac12, 0, \frac12, 1,\frac32)$ at $t=20$.}
\label{f:x3s}
\end{figure}

Next we consider the asymptotics for $y\to-\infty$.
Starting with the non-pivot column $g_1=2$,
the only column to its left is $i=1$.
Then, we have $Y_{12}=\emptyset$, and
$\rank(Y_{12}|1)= \rank(Y_{12}|2)= \rank(Y_{12}|1,2)=1$.
Consequently, the pair $[1,2]$ identifies an asymptotic line soliton
as $y\to-\infty$.
For $g_2=4$ we consider $i=1,2,3$ and find that
the rank conditions are satisfied only for $i=2$. In this case,
$Y_{24}= \bigl(\begin{smallmatrix}0\\1\end{smallmatrix}\bigr)$,
so $\rank(Y_{24})= 1=N-1$ and
$\rank(Y_{24}|2)= \rank(Y_{24}|4)= \rank(Y_{24}|2,4)=2$.
Hence $[2,4]$ is the unique asymptotic line soliton as $y\to-\infty$
associated to the non-pivot column $g_2=4$.
In a similar way, we can uniquely identify the last asymptotic line soliton
as $y\to-\infty$ as given by the indices $[3,5]$.
To summarize, there are $N_+=2$ outgoing line solitons
given by the index pairs $[1,4]$ and $[3,5]$,
and there are $N_-=3$ incoming line solitons
given by the index pairs $[1,2]$, $[2,4]$ and $[3,5]$.
A snapshot of the solution at $t=-32$ is shown in Fig.~\ref{f:x3s}.

\noindent
{\em Example 2}:\, Next, consider the $\tau$-function with $N=3$ and $M=6$ 
generated by the coefficient matrix in RREF
\begin{equation*}
A= \begin{pmatrix}
1 &1 &1 &0 &0 &0\\
0 &0 &0 &1 &0 &\!-1\\
0 &0 &0 &0 &1 &2
\end{pmatrix}\,.
\label{e:inelastic3s}
\end{equation*}
Again, we first determine the asymptotic line solitons as $y\to\infty$.
In this case, the pivot columns of~$A$ are labeled by 
$\{e_1,e_2,e_3\}=\{1,4,5\}$.
So, the asymptotic line solitons as $y\to\infty$
are given by the index pairs $[1,j_1]$, $[4,j_2]$ and $[5,j_3]$ for
some $j_1,\, j_2,\,j_3$.
Starting with the first pivot, $e_1=1$, we take $j=2,3,\dots$
and check the rank of the sub-matrix $X_{ij}$ in each case.
When $j=2$ we have
$X_{12}= \left(\!
\begin{smallmatrix}
1 &0 &0 &0\\ 0 &1 &0 &\!-1\\ 0 &0 &1 &2
\end{smallmatrix}\!\right)\,$,
so $\rank(X_{12})=3>N-1$.  Hence, according to 
Proposition\ref{P:pairing}(ii),
the index pair $[1,2]$ does not correspond to an asymptotic
line soliton as $y\to\infty$. We then take $j=3$ and consider
the sub-matrix
$X_{13}= \left(\!\begin{smallmatrix}
1 &0 &0\\ 0 &0 &\!-1\\ 0 &1 &2
\end{smallmatrix}\!\right)\,$.
Since $\rank(X_{13})=2$ and $\rank(X_{13}|1)= \rank(X_{13}|3)=
\rank(X_{13}|1,3)= 3$, the rank conditions in Proposition\ref{P:pairing}(ii)
are satisfied.
Therefore the index pair $[1,3]$ corresponds to an asymptotic line soliton
as $y\to\infty$. Moreover, by considering $j=4,5,6$ one can easily check that the
rank conditions are no longer satisfied.
Thus $[1,3]$ is the \textit{unique} asymptotic line soliton associated
with the pivot index $e_1=1$ as $y\to\infty$.
Proceeding in a similar way, we find that for the pivot column $e_2=4$,
the rank conditions are only satisfied when $j=5$, since
$X_{45}= \left(\!\begin{smallmatrix}
1 &1 &1 &0\\ 0 &0 &0 &\!-1\\ 0 &0 &0 &2
\end{smallmatrix}\!\right)\,$,
is of rank 2, and $\rank(X_{45}|4)= \rank(X_{45}|5)=
\rank(X_{45}|4,5)= 3$.
Therefore, the index pair $[4,5]$ corresponds to an asymptotic line soliton
as $y\to\infty$. Finally, we find that the third asymptotic
line soliton as $y\to\infty$ is given by the index pair $[5,6]$.
                                                                                
Next, we proceed to determine the asymptotic line solitons as $y\to-\infty$.
The non-pivot columns of $A$ are labeled by the indices
$g_1=2$ $g_2=3$ and $g_3=6$.
For $g_1=2$, the only possible value of $i<j$ is $i=1$.
In this case $Y_{12}=\emptyset$, so $\rank(Y_{12})=0$ and
$\rank(Y_{12}|1)= \rank(Y_{12}|2)= \rank(Y_{12}|1,2)= 1$.
Thus, the pair $[1,2]$ identifies an asymptotic line soliton as 
$y\to-\infty$. For $g_2=3$ we consider $i=2,1$.
When $i=2$, the rank conditions in Proposition~\ref{P:pairing}(ii)
are satisfied,
leading to the asymptotic line soliton $[2,3]$ as $y\to-\infty$.
Similarly, it is easy to verify that for $g_3=6$ the index pair $[4,6]$
uniquely identifies the asymptotic line soliton as $y\to-\infty$.
Summarizing, there are $N_+=3$ asymptotic line solitons as $y\to\infty$
identified by the index pairs $[1,3]$, $[4,5]$ and $[5,6]$,
and there are $N_-=3$ asymptotic line solitons as $y\to-\infty$
identified by the index pairs $[1,2]$, $[2,3]$ and $[4,6]$.
A snapshot of this $(3,3)$-soliton solution
at $t=-20$ is shown in Fig.~\ref{f:x3s}b.

So far we have discussed the properties of the generic
$(N_-,N_+)$-soliton solutions of KPII, and have shown how to characterize
the asymptotic line solitons as $y \to \pm \infty$ from the corresponding 
$\tau$-function. In the next section, we investigate an important subclass 
of the $(N_-,N_+)$-soliton solutions called the {\em elastic} 
$N$-soliton solutions.

\section{Elastic $N$-soliton solutions}
\label{s:elastic}
\noindent
We begin this section by introducing the notion of equivalence classes of 
soliton solutions and their $\tau$-functions, both of which will
play important roles in this section.
\begin{definition}
(Equivalence class)~
Let $\Theta$ denote the set of all exponential phase combinations 
whose coefficients are non-zero in the $\tau$-function of 
Eq.~\eqref{e:tauexp}. Two $tau$-functions are said to be in the same 
equivalence class if they contain the same set $\Theta$ (up to an overall 
exponential phase factor). All $(N_-,N_+)$-soliton solutions
of KPII generated by an equivalence class of $\tau$-functions form an equivalence 
class of solutions.
\label{D:equivalenceclass}
\end{definition}
It is clear from the above definition that the $\tau$-functions 
in a given equivalence class can be viewed as positive-definite sums
of the \textit{same} exponential phase combinations but with
different sets of coefficients. Such $\tau$-functions 
are parametrized by the same set of phase parameters $k_1,\ldots,k_M$, 
but the constants $\theta_{m0}$ in the phase $\theta_m$ are different.
Moreover, the irreducible coefficient matrices associated with
the $\tau$-functions have exactly the same sets of vanishing and 
non-vanishing minors, but the magnitudes of the non-vanishing minors 
are different for different matrices.
Then, an important consequence of Proposition \ref{P:soliton} 
is that for all $(N_-,N_+)$-soliton solutions 
in the same equivalence class, the corresponding asymptotic 
line solitons arise from the {\em same} $i \to j$ single-phase transition, 
and are therefore labeled by the same 
index pair $[i,j]$. Proposition \ref{P:pairing}
then implies that the coefficient matrices associated with
the $\tau$-functions in the same equivalence class have identical
sets of pivot and non-pivot indices which identify respectively, the asymptotic
line solitons as $y \to \infty$ and as $y \to -\infty$.
Thus, solutions in the same equivalence class can differ only  
in the position of each asymptotic line 
solitons and in the location of each interaction vertex. As a result,
any $(N_-,N_+)$-soliton solution of KPII can be transformed into 
any other solution in the same equivalence class by spatio-temporal 
translations of the individual asymptotic line solitons.

The KPII equation~\eqref{e:KP}
is invariant under the symmetry $(x,y,t) \to (-x,-y,-t)$.
Thus, if $u(x,y,t)$ is an $(M-N,N)$-soliton solution of KPII,
then $u(-x,-y,-t)$ is also an $(N_-,N_+)$-soliton solution
whose incoming and outgoing line solitons are reversed so that
$N_-=N$ and $N_+=M-N$. In general,
$u(x,y,t)$ and $u(-x,-y,-t)$ belong to two different equivalence
classes of solutions, and so do their generating $\tau$-functions.  
However, the function $\tau(-x,-y,-t)$ generating the solution $u(-x,-y,-t)$ 
is not by itself a $\tau$-function according to Eq.~\eqref{e:tauexp}.
Using Eq.~\eqref{e:tauexp}, $\tau(-x,-y,-t)$ can be expressed as
$\tau(-x,-y,-t)= e^{-\theta_{1,\dots,M}}\, \tau'(x,y,t)\,,$
where the function 
\begin{equation*}
\tau'(x,y,t)=  \!\!\!
  \sum\limits_{1\le m_1<\dots<m_N\le M} \!\!\!
   A(m_1,\dots,m_N)\,\,
    \exp[\theta(l_1,\dots,l_{M-N})]\!\!\!
\prod_{1\le s < r\le N}(k_{m_{r}}-k_{m_{s}})
\end{equation*}
is a positive definite sum of exponential phase combinations 
labeled by the set of indices $\{l_1,\ldots,l_{M-N}\}$, which is the 
complement of $\{m_1,\ldots,m_N\}$ in $\{1,2,\ldots,M\}$. 
Moreover, since $\tau'(x,y,t)$ only differs from $\tau(-x,-y,-t)$
by an overall exponential phase factor, it should be clear from 
Eq.~\eqref{e:u} that they both generate the same solution 
$u(-x,-y,-t)$. The correspondence between equivalence classes
of solutions and their $\tau$-functions related via the symmetry 
$(x,y,t) \to (-x,-y,-t)$ leads to the following notion of {\em duality}. 
\begin{definition}
(Duality)~
Two equivalence classes of $\tau$-functions are said to be 
dual if they are parametrized by the same set of phase parameters
$\{k_1,k_2,\ldots,k_M\}$, but correspond to complementary sets
of exponential phase combinations $\Theta$ and $\Theta'$. That is,
$$ \theta(m_1,\ldots,m_N) \in \Theta \,\, \Leftrightarrow
\,\, \theta(l_1,\ldots,l_{M-N}) \in \Theta'\,,$$
where the sets $\{m_1,\ldots,m_N\}$
and $\{l_1,\ldots,l_{M-N}\}$ form a disjoint partition of the
integers $\{1,2,\ldots,M\}$. Similarly, two equivalence classes of 
$(N_-,N_+)$-soliton solutions are dual if they are generated by 
dual equivalence classes of $\tau$-functions.
\label{D:duality}
\end{definition}
\noindent
In particular, if a given $(M-N,N)$-soliton solution $u(x,y,t)$
belongs to a certain equivalence class, then the corresponding
$(N,M-N)$-soliton solution $u(-x,-y,-t)$ belongs to the dual 
equivalence class.

An interesting subclass of $(N_-,N_+)$-soliton solutions 
are the elastic $N$-soliton solutions of KPII as mentioned in 
section~\ref{s:introduction}. These can be defined as follows.
\begin{definition}
A $(N_-,N_+)$-soliton solution is called elastic if it belongs
to an equivalence class which is its own dual.
\label{D:elastic}
\end{definition}
Clearly, in this case we have $N_+=N_-=N$ and $M=2N$.
Moreover, the amplitudes and directions of the $N$ incoming line solitons
coincide with those of the $N$ outgoing line solitons.
Thus, an elastic $N$-soliton solution is generated by a ``self-dual'' 
$\tau$-function which is a positive 
definite sum over a set $\Theta$ of exponential phase combinations such 
that the following condition holds:
\begin{equation*}
\theta(m_1,\ldots,m_N) \in \Theta \,\, \Leftrightarrow 
\,\, \theta(l_1,\ldots,l_N) \in \Theta \,, \quad \forall \,
\{m_1,\ldots,m_N\} \sqcup \{l_1,\ldots,l_N\} = \{1,2,\ldots,2N\}\,.  
\end{equation*}
We outline below the main properties of the elastic $N$-soliton solutions.
Additional details regarding these solutions can be found in
Refs.~\cite{GBSCYK,Kodama}.
\begin{proposition} The elastic $N$-soliton solutions of KPII are 
characterized as follows.
\begin{enumerate}
\item
Each elastic $N$-soliton solution $u(x,y,t)$ and the dual solution
$u(-x,-y,-t)$ belong to the same equivalence class.
\item
The $\tau$-function corresponding an elastic $N$-soliton solution
has $M=2N$ distinct phase parameters and an $N \times 2N$, irreducible,  
rank $N$ coefficient matrix $A$ whose $N \times N$ minors satisfy the
duality conditions:
\begin{equation}
A(m_1,\ldots,m_N) = 0 \,\, \Leftrightarrow \,\, 
A(l_1,\ldots,l_N) = 0\,,
\label{e:dualminor}
\end{equation}
where the indices $\{m_1,\ldots,m_N\}$ and
$\{l_1,\ldots,l_N\}$ form a disjoint partition of integers
$\{1,2,\ldots, 2N\}$.
\item
Each elastic $N$-soliton solution exactly $N$ 
asymptotic line solitons as $y \to \pm \infty$ identified by the
same index pairs $[e_n,g_n]$ with $e_n < g_n, \, n=1,\ldots, N$. 
The indices $e_1, e_2, \ldots, e_N$ and $g_1, g_2, \ldots, g_N$ label 
respectively, the pivot and non-pivot columns of the coefficient matrix 
$A$. Hence, they form a disjoint partition of integers $\{1,2,\ldots, 2N\}$.
\item
The amplitude and direction of the $n^{\mathrm {th}}$ asymptotic line 
soliton $[e_n,g_n]$ are the same as $y \to \pm \infty$, and
are given in terms of the phase parameters as $a_n = k_{g_n} - k_{e_n}$ 
and $c_n = k_{g_n} + k_{e_n}$.
\end{enumerate} 
\label{P:elastic}
\end{proposition}

The set $S_N := \{(a_n,c_n)|a_n > 0\}_{n=1}^N \subset \Real^{2N}$ 
of all {\em admissible} $N$-tuples of amplitude-direction pair 
associated with elastic $N$-soliton solution will be called the 
{\em soliton parameter space}. An element $\{a_n,c_n\}_{n=1}^N \in S_N$ 
of soliton parameters is admissible, if it yields a set of $2N$ 
{\em distinct} phase parameters 
$K_N := \{k_n^\pm \}_{n=1}^N$ where $k_n^\pm = (c_n \pm a_n)/2$
and $k_n^- < k_n^+$.
By sorting the elements of $K_N$ in increasing order 
$k_1 < \ldots < k_{2N}$, one obtains the ordered set 
$K = \{k_1, k_2, \ldots, k_{2N}\}$ of phase
parameters associated with the $\tau$-function. The positions
of the phase parameters $k_n^\pm$ of the $n^{\mathrm {th}}$ line
soliton can be labeled uniquely 
within the ordered set $K$ by an ordered pair 
of indices $[i_n,j_n]$ such that $i_n < j_n$. That is, 
$k_n^- = k_{i_n}, \, k_n^+ = k_{j_n}$. Since $[i_n,j_n]$ also
identify two distinct columns of the coefficient matrix $A$, it
follows that $i_n = g_n$ which labels a pivot column, and $j_n = g_n$
which labels a non-pivot column of $A$, for elastic $N$-soliton solutions.
We refer to this 
identification between the sets $K_N$ and $K$ as {\em phase pairing}
which defines a map $S_N \rightarrow S$, where $S$ is the set of all 
possible choices of $N$ distinct integer pairs $\{[i_n,j_n]\}_{n=1}^N$ from
$\{1,2,\ldots,2N\}$. This map identifies each set of soliton parameters 
$\{(a_n,c_n)\}_{n=1}^N \in S_N$ to a set 
$\{[e_n,g_n]\}_{n=1}^N \in S$ of (pivot, non-pivot) index pairs.
Note however that distinct elements of $S_N$ can in fact lead to the 
\textit{same pairing} if the elements of the corresponding sets $K_N$
are ordered in exactly identical fashion.
Thus, phase pairing induces a partition of the soliton parameter
space~$S_N$ into several disjoint sectors. Each sector is distinguished
by a single element $\{[e_n,g_n]\}_{n=1}^N \in S$ of distinct integer pairs
which labels the $N$ asymptotic line solitons corresponding to 
any set $\{(a_n,c_n)\}_{n=1}^N$ of soliton parameters chosen from
that sector in $S_N$. Therefore, the total number of such disjoint sectors 
is given by the number of elements of the set $S$ namely, $|S| = (2N-1)!!$.
Furthermore, given any set of soliton parameters $\{(a_n,c_n)\}_{n=1}^N$
from one of these sectors in $S_N$, it is possible to construct
a coefficient matrix~$A$ in RREF satisfying Conditions~\ref{c:positive}
and \ref{c:irreducible} and whose pivot and non-pivot columns
are labeled by the corresponding set of index pairs
$\{[e_n,g_n]\}_{n=1}^N \in S$ obtained via phase pairing.
The matrix $A$ is constructed by using the rank conditions
of Proposition~\ref{P:pairing}(ii) and the duality condition
Eq.~\eqref{e:dualminor} in
Proposition~\ref{P:elastic}(ii). However, it is not unique but is 
determined up to some free parameters. Therefore, the matrix $A$ obtained 
in this way together with the set~$K$ of phase parameters, produce 
an equivalence class of $\tau$-functions from \eqref{e:tauexp}. The directions 
and amplitudes of the asymptotic 
line solitons in the corresponding equivalence class of elastic $N$-soliton
solutions coincide with the set of soliton parameters 
$\{(a_n,c_n)\}_{n=1}^N \in S_N$ that was originally chosen. We illustrate
these facts in the following example where we explicitly construct an elastic 
$3$-soliton solution only from its soliton parameters.

\noindent
{\em Example}:\, We start with the set 
$\{(a_1,c_1), (a_2,c_2), (a_3,c_3)\}=\{(5/2,-5/2), (5/2, -1/2), (7/4,5/4)\} \in S_3$.
From this we construct the set of {\em unordered} phase parameters
$K_3 := \{k_n^\pm = (c_n \pm a_n)/2\}_{n=1}^3$ whose elements are 
$k_1^+=0,\,k_1^-=-5/2,\, k_2^+=1,\,k_2^-=-3/2,\,k_3^+=3/2,\,k_3^-=-1/4$.
Note that the elements of $K_3$ are all distinct. Sorting these elements
of $K_3$ in increasing order, we obtain the ordered set of phase parameters
$K = \{-5/2, -3/2, -1/4, 0, 1, 3/2\}$. Comparing the elements of the sets
$K_3$ and $K$, we obtain the following phase pairing:\,
$k_1^-=k_1,\,k_1^+=k_4,\,k_2^-=k_2,\,k_2^+=k_6,\,k_3^- = k_3,\,k_3^+= k_5$. 
Hence, the asymptotic line solitons are
labeled by the index pairs $[1,4]$, $[2,6]$ and $[3,5]$, where
$\{e_1,e_2,e_3\}=\{1,2,3\}$ are the pivot indices and 
$\{g_1,g_2,g_3\} =\{4,6,5\}$ are the non-pivot indices of the
corresponding coefficient matrix $A$. Note also that
the pivot indices are sorted but the non-pivot indices are unsorted.
Next, we outline the construction of the coefficient matrix $A$ in RREF
satisfying Conditions~\ref{c:positive} and \ref{c:irreducible}, and whose 
pivot and non-pivot column indices are specified above. The construction 
proceeds in several steps.
 
\noindent\textit{Step~1.}~
We start with a $3\times6$ matrix in RREF with pivot and non-pivot
columns labeled respectively, by the indices $\{1,2,3\}$ and $\{4,6,5\}$:
\begin{equation*}
A= \begin{pmatrix}
 1 &0 &0 &u_1 &v_1 &w_1 \\
 0 &1 &0 &\!-u_2 &\!-v_2 &\!-w_2 \\
 0 &0 &1 &u_3 &v_3 &w_3 \\
\end{pmatrix}\,,
\end{equation*}
where $\{u_l, v_l, w_l\}_{l=1}^3$ are nonnegative numbers.
The negative signs in the second row arise due to 
Condition~\ref{c:positive}(c) which demands that if any given $3 \times 3$ 
minor of $A$ is non-zero, then it must be positive. 

\begin{figure}[t!] 
\sfigs{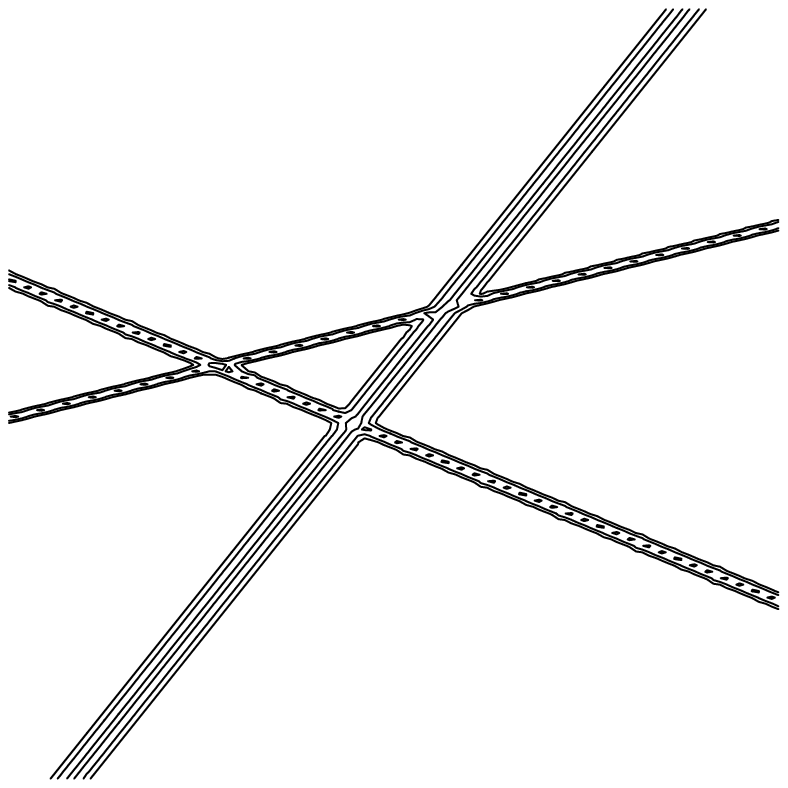}{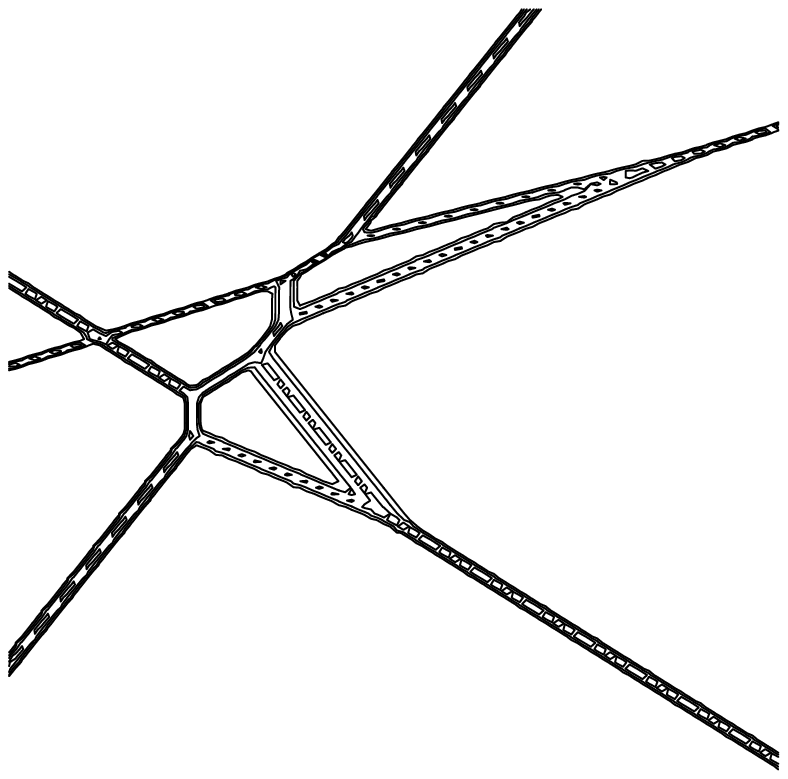}{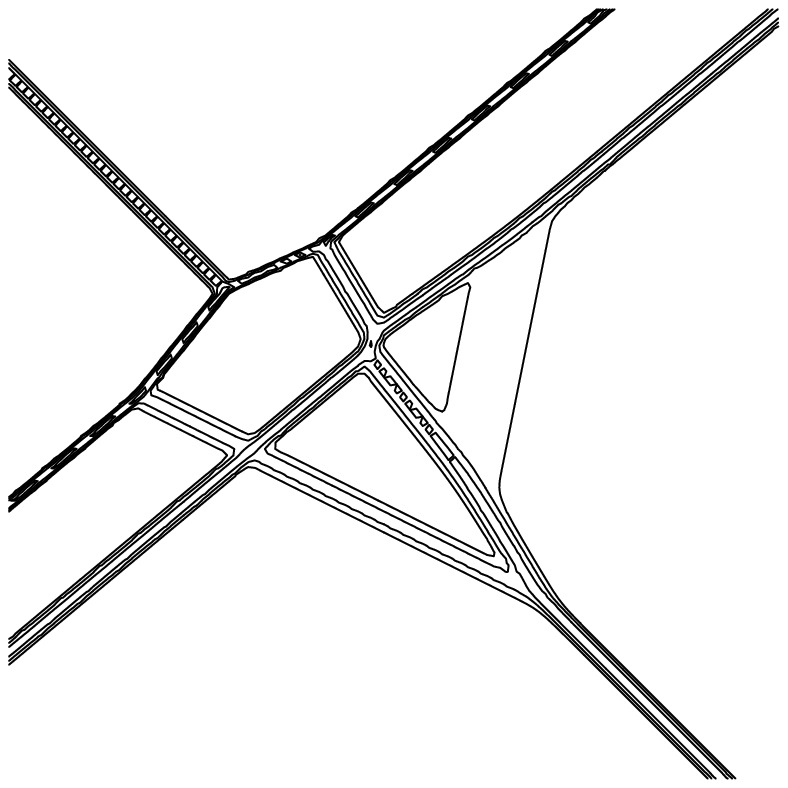}{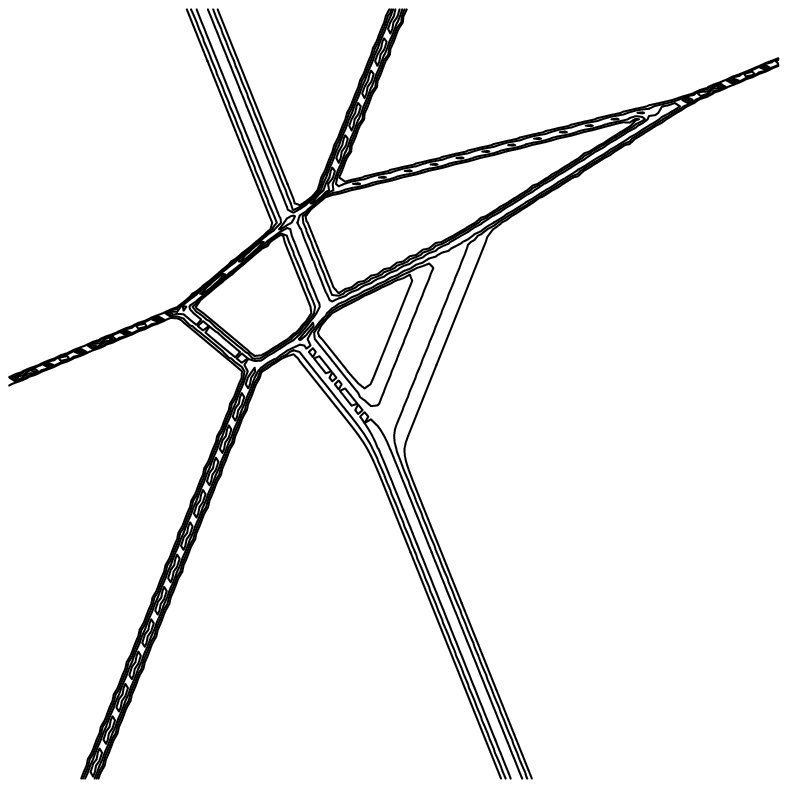}
\stext{(a)\,\, [1,2], [3,4], [5,6]}{(b)\,\,[1,3], [2,5], [4,6]}
{(c)\,\,[1,4], [2,6], [3,5]}{(d)\,\, [1,5], [2,4], [3,6]}
\kern\medskipamount
\caption{Examples of elastic 3-soliton solutions.}
\label{f:3s}
\kern-\bigskipamount
\end{figure}

\noindent\textit{Step 2.}~
In order to obtain further information about $A$, we
apply the rank conditions in Proposition~\ref{P:pairing}(ii) to
the sub-matrices $X_{ij}$ and $Y_{ij}$ associated with each line
soliton $[i,j]$. For example, starting with the line soliton $[1,4]$
as $y \to -\infty$ and by considering the sub-matrix
$Y_{14}= \bigl(\begin{smallmatrix} 0 &0\\ 1 &0 \\ 0 &1 \\
\end{smallmatrix}\bigr)\,$, we find that $\rank(Y_{14})=2$.
Then, $\rank(Y_{14}|4)$ must be 3, which means that the minor
$A(2,3,4)=u_1 \neq 0$. Now suppose $v_1 = 0$, then the non-negativity of
the minors in Condition~\ref{c:positive}(c) implies that
$A(2,4,5) = -u_1v_3 \geq 0$ and $A(3,4,5) = -u_1v_2 \geq 0$, 
whose only solution is $v_2=v_3=0$ since $u_1 \neq 0$. But then the
$5^{\mathrm {th}}$ column of $A$ contains no nonzero elements, violating
the irreducibility of $A$ in Condition~\ref{c:irreducible}. Thus, 
$v_1 \neq 0$, and similar arguments lead to $w_1 \neq 0$. Then from 
$A(l,\alpha, \beta) \geq 0$ where $l \in \{2,3\}$ and 
$\alpha, \beta \in \{4,5,6\}$, we can deduce that if $u_l = 0$, then 
$v_l=w_l=0$ for $l=2,3$.  Consequently, the only
nonzero element in each of the $2^{\mathrm {nd}}$ and $3^{\mathrm {rd}}$  
row of $A$ would be the pivot entry, and again this would violate 
Condition~\ref{c:irreducible}. Hence, we must also have $u_2 \neq 0$ and 
$u_3 \neq 0$. As our goal is to obtain only one representative 
matrix $A$ associated to the equivalence class of $\tau$-function, we
simplify subsequent calculations by choosing a particular normalization 
such that nonzero elements $u_1=v_1=w_1=1$. Then we have
\begin{equation*}
A= \begin{pmatrix}
 1 &0 &0 &1 &1 &1 \\
 0 &1 &0 &\!-u_2 &\!-v_2 &\!-w_2 \\
 0 &0 &1 &u_3 &v_3 &w_3 \\
\end{pmatrix}\,.
\end{equation*}
\noindent\textit{Step 3.}~
Next, we consider the line soliton $[3,5]$ as $y \to \infty$
and the associated sub-matrix  
$X_{35}= \bigl(\begin{smallmatrix} 1 &0 &1\\ 0 &1 &\!-w_2\\0 & 0& w_3 \\
\end{smallmatrix}\bigr)\,$. From the condition $\rank(X_{35}) \leq N-1=2$, 
we have $\det(X_{35}) = A(1,2,6) = 0$, which implies that $w_3=0$.
Moreover, since the minor $A(1,2,6) = 0$, it follows from the duality 
condition Eq.~\eqref{e:dualminor} that $A(3,4,5)=u_2-v_2=0$. 
Hence $v_2=u_2 \neq 0$. 
Applying the duality condition again to the minor consisting of the pivot
columns, we obtain $A(1,2,3)=1 \neq 0 \Rightarrow A(4,5,6) \neq 0$.
In particular, this means that the $5^{\mathrm {th}}$ and $6^{\mathrm {th}}$
columns of $A$ are linearly independent, and that the
sub-matrix $X_{14}= \bigl(\begin{smallmatrix} 1 &1\\\!-v_2 &\!-w_2\\ 
v_3 &w_3 \\ \end{smallmatrix}\bigr)\,$ associated with the $[1,4]$ line soliton
as $y \to \infty$, has rank 2. Then it follows from the rank conditions in 
Proposition~\ref{P:pairing}(ii) that 
$\rank(X_{14}|1)=3 \Rightarrow A(1,5,6) = v_3\,w_2 \neq 0$. Thus, we have
$v_3 \neq 0$ and $w_2 \neq 0$. Finally, imposing the non-negativity
condition on the remaining minors, we obtain the following form of 
the coefficient matrix $A$
\begin{equation*}
A= \begin{pmatrix}
 1 &0 &0 &1 &1 &1 \\
 0 &1 &0 &\!-u_2 &\!-u_2 &\!-w_2 \\
 0 &0 &1 &u_3 &v_3 & 0\\
\end{pmatrix}\,.
\end{equation*}
where the remaining free parameters satisfy $0 < w_2 < u_2$ and 
$0 < v_3 < u_3$.
Thus, starting only from the soliton parameters,
we have constructed a 4-parameter family of coefficient matrices
corresponding to an equivalence class of elastic $3$-soliton solutions
whose asymptotic line solitons are labeled by the index pairs
$[1,4],[2,5],[3,6]$. An elastic $3$-soliton solution generated by the 
above coefficient matrix $A$ with $(u_2,w_2,u_3,v_3)=(1,2/3,2/3,3/5)$
and $K = \{-5/2, -3/2, -1/4, 0, 1, 3/2\}$ (as above), 
is shown in Fig.~\ref{f:3s}(c).

By further investigating the combinatorial properties
of the coefficient matrix $A$, it is possible to obtain 
additional information regarding the classification scheme
for the elastic $N$-soliton configuration space.
These results are presented below.
\begin{proposition}
Each elastic $N$-soliton configuration is described by a set 
$\{[e_n,g_n]\}_{n=1}^N$ of distinct integer pairs with 
$e_n < g_n,\, n=1,\ldots,N$. The indices $e_n$ label the pivot columns 
and the indices $g_n$ label the non-pivot columns of the irreducible 
coefficient matrix $A$ in RREF. In addition, the following results hold:
\begin{enumerate}
\item
Without any loss of generality, the pivot indices can be 
ordered as $1=e_1<e_2< \ldots < e_N < 2N$.
Each pivot index $e_n \, n=1,\ldots,N$ satisfies the inequality
$ n \leq e_n \leq 2n-1$.
Moreover, the number of possible ways of choosing the pivots 
are given by the Catalan number
$C_N = \displaystyle \frac{(2N)!}{N!(N+1)!}$.
\item
For a fixed ordered set $\{e_1, \ldots, e_N\}$
of pivots, the number of possible choices of a (unordered) set
$\{g_1, \ldots, g_N \}$ of non-pivot indices such that 
$e_n < g_n, \, \forall\, n=1,\ldots,N$, is
given by $\displaystyle \prod_{n=1}^N(2n-e_n)$.
\item
The total number of ways of choosing $N$ distinct pairs in
the set $S$ is given by $(2N-1)!!$.
\end{enumerate}
\label{P:elasticsectors}
\end{proposition}

Note that the requirement that the set of integer pairs 
$\{[e_n,g_n]\}_{n=1}^N$ be distinct was already stated (without proof)
in Ref.~\cite{Kodama}, and some of the above-listed consequences
were also obtained there.

We illustrate the results in Proposition~\ref{P:elasticsectors} by presenting
the classification scheme for the elastic $3$-soliton solution space.
This is achieved by enumerating all possible arrangements of the
pivot positions in the irreducible coefficient matrix $A$ in RREF.
In this case, $N=3$ and $A$ is a $3 \times 6$ matrix with 3 pivots whose
possible (column) positions are determined by Proposition~\ref{P:elasticsectors}(i) 
as follows:\, $e_1 =1, \, 2 \leq e_2 \leq 3, \, 3 \leq e_3 \leq 5$.
Thus, the total number of pivot configurations is given by 
$C_3 = 6!/(3!\,4!) = 5$. Thus, this classification
scheme gives rise to 5 subclasses
of elastic $3$-soliton solutions. The number of inequivalent types of
solutions in each subclass is determined by all possible $\{[e_n,g_n]\}_{n=1}^3$ 
pairings for a given choice of the pivot positions $\{e_1,e_2,e_3\}$.
These are obtained from Proposition~\ref{P:elasticsectors}(ii), and are
itemized below.
\begin{enumerate}
\vspace{-0.15 in}
\item[(i)]\, Pivot positions:\, $\{e_1,e_2,e_3\} = \{1,2,3\}$.
Total number of distinct pairings = $\displaystyle \prod_{n=1}^3(2n-e_n) = 3! = 6$.
List of inequivalent elastic $3$-soliton solutions:
\begin{gather*}
\{[1,4],~[2,5],~[3,6]\}\,,\quad \{[1,5],~[2,4],~[3,6]\}\,,\quad
\{[1,6],~[2,4],~[3,5]\}\,,
\\
\{[1,4],~[2,6],~[3,5]\}\,,\quad \{[1,5],~[2,6],~[3,5]\}\,, \quad
\{[1,6],~[2,5],~[3,4]\}\,.
\end{gather*}
\vspace{-0.25in}
\item[(ii)]\, Pivot positions:\, $\{e_1,e_2,e_3\} = \{1,2,4\}$.
Total number of distinct pairings = 4.
List of inequivalent elastic $3$-soliton solutions:
\begin{gather*}
\{[1,3],~[2,5],~[4,6]\}\,,\quad \{[1,3],~[2,6],~[4,5]\}\,,\quad
\{[1,5],~[2,3],~[4,6]\}\,, \quad \{[1,6],~[2,3],~[4,5]\}\,.
\end{gather*}
\vspace{-0.25in}
\item[(iii)]\, Pivot positions:\, $\{e_1,e_2,e_3\} = \{1,2,5\}$.
Total number of distinct pairings = 2.
List of inequivalent elastic $3$-soliton solutions:\,
$\{[1,3],~[2,4],~[5,6]\}\,,\quad \{[1,4],~[2,3],~[5,6]\}$.
\item[(iv)]\, Pivot positions:\, $\{e_1,e_2,e_3\} = \{1,3,4\}$.
Total number of distinct pairings = 2.
List of inequivalent elastic $3$-soliton solutions: \,
$\{[1,2],~[3,5],~[4,6]\}\,,\quad \{[1,2],~[3,6],~[4,5]\}$.
\item[(v)]\, Pivot positions:\, $\{e_1,e_2,e_3\} = \{1,3,5\}$.
Total number of distinct pairings = 1.
Elastic $3$-soliton solution:\, $\{[1,2],~[3,4],~[5,6]\}$.
\end{enumerate}
Thus, the total number of inequivalent $3$-soliton solutions is
$6+4+2+2+1 = 15 = 5!!$ as given by Proposition~\ref{P:elasticsectors}(iii).
Fig.~\ref{f:3s} shows a sample from the fifteen inequivalent cases.
Fig.~\ref{f:3s}(a) shows the previously known {\em ordinary} $3$-soliton 
solution (cf. section \ref{s:introduction}), while the remaining solutions
are new, and they exhibit resonant interactions.

\section{Conclusion}
\label{s:conclusion}

\noindent
In this article, a family of multi-soliton solutions
of the KPII equation have been studied. These solutions are generated 
by $\tau$-functions 
which are expressed as positive definite linear combinations of 
exponential phases that are linear in the variables $x,y,t$.
It is remarkable that such a simple form of the $\tau$-function generates
multi-soliton configurations which exhibit a rich variety of time dependent 
spatial structures including resonant interactions and web patterns.
The asymptotic analysis of the $tau$-function in the $xy$-plane
reveals that the solution decays exponentially except along certain
directions which are characterized by the transition between two dominant 
exponential phase combinations which have all {\em but one}  
phases in common. All such ("non-decaying") directions for any given
solution can be explicitly identified by analyzing the 
$N \times M$ matrix coefficient $A$ associated with the $\tau$-function.
In particular, as $y \to \infty$ there are $N$ directions which
can be identified with the pivot columns of $A$; while as $y \to -\infty$, 
there are $M-N$ directions which can be identified with the non-pivot 
columns of $A$.

When $M=2N$, the general line soliton solutions contain the special subclass 
of the elastic $N$-soliton solutions. Each elastic $N$-soliton solution 
has a set of $N$ directions in the $xy$-plane as $y \to \infty$, and
an identical set of $N$ directions as $y \to -\infty$ along which the 
solution does not decay. Moreover, for any given $N$, the solution space of 
elastic $N$-solitons can be decomposed into $(2N-1)!!$ distinct regions, 
each region corresponding to inequivalent types of solutions. 
It is interesting to note that the previously known ordinary 
$N$-solitons form only one of these types.
Thus, the space of elastic $N$-soliton
solutions of KPII appears to be much richer than previously thought.

It is significant that solutions exhibiting
similar features of soliton resonance and web structure have also been
obtained in several other (2+1)-dimensional integrable systems,
besides KPII. These solutions were also derived by direct
algebraic methods similar to the approach taken here.
Therefore, it is reasonable to expect that the results developed in this 
work for KPII will also be useful to characterize soliton solutions
in a variety of other (2+1)-dimensional integrable systems.

\section*{Acknowledgments}
We thank Yuji Kodama, Mark Ablowitz and Dmitry Pelinovsky for many valuable 
discussions. This work was partially supported by the National Science Foundation 
under grant numbers DMS-0307181 and DMS-0101476.

\catcode`\@ 11
\def\journal#1&#2,#3 (#4){\begingroup \let\journal=\d@mmyjournal {\frenchspacing\sl #1\/\unskip\,} {\bf\ignorespaces #2}\rm, #3 (#4)\endgroup}
\def\d@mmyjournal{\errmessage{Reference foul up: nested \journal macros}}
\def\title#1{{``#1''}}
\def\@biblabel#1{#1.}
\catcode`\@ 12

\end{document}